\documentclass[preprint,12pt]{elsarticle}

\usepackage{amssymb}
\usepackage{subfigure}
\usepackage{float}
\usepackage{color}

\journal{Nuclear Physics A}

\begin{document}

\begin{frontmatter}



\title{Performance of the Silicon-On-Insulator Pixel Sensor for X-ray Astronomy, XRPIX6E, Equipped with Pinned Depleted Diode Structure}

\author[kyoto]{Sodai~Harada}
\ead{harada.sodai.78s@kyoto-u.jp}
\author[kyoto]{Takeshi~Go~Tsuru} \author[kyoto]{Takaaki~Tanaka} \author[kyoto]{Hiroyuki~Uchida} \author[kyoto]{Hideaki~Matsumura} 
\author[kyoto]{Katsuhiro~Tachibana} \author[kyoto]{Hideki~Hayashi}
\author[miyazaki]{Ayaki~Takeda} \author[miyazaki]{Koji~Mori} \author[miyazaki]{Yusuke~Nishioka} \author[miyazaki]{Nobuaki~Takebayashi}
\author[miyazaki]{Shoma~Yokoyama} \author[miyazaki]{Kohei~Fukuda}
\author[kek]{Yasuo~Arai} \author[kek]{Ikuo~Kurachi}
\author[shizuoka]{Shoji~Kawahito} \author[shizuoka]{Keiichiro~Kagawa} \author[shizuoka]{Keita~Yasutomi} \author[shizuoka]{Sumeet~Shrestha} \author[shizuoka]{Syunta~Nakanishi}
\author[okinawa]{Hiroki~Kamehama}
\author[tokyorika]{Takayoshi~Kohmura} \author[tokyorika]{Kouichi~Hagino} \author[tokyorika]{Kousuke~Negishi} \author[tokyorika]{Kenji~Oono} \author[tokyorika]{Keigo~Yarita}

\address[kyoto]{Department of Physics, Graduate School of Science, Kyoto University, Kitashirakawa Oiwake-cho, Sakyo-ku, Kyoto 606-8502, Japan}
\address[miyazaki]{Department of Applied Physics, Faculty of Engineering, University of Miyazaki,1-1 Gakuen Kibana-dai Nishi, Miyazaki 889-2192, Japan}
\address[kek]{Institute of Particle and Nuclear Studies, High Energy Accelerator Research Org., KEK, 1-1 Oho, Tsukuba 305-0801, Japan}
\address[shizuoka]{Research Institute of Electronics, Shizuoka University, Johoku 3-5-1, Naka-ku, Hamamatsuu, Shizuoka 432-8011, Japan}
\address[okinawa]{National Institute of Technology, Okinawa College, Henoko 905, Nago-shi, Okinawa 905-2192, Japan}
\address[tokyorika]{Department of Physics, Faculty of Science and Technology, Tokyo University of Science, 2641 Yamazaki, Noda, Chiba 278-8510, Japan}

\begin{abstract}
We have been developing event driven X-ray Silicon-On-Insulator (SOI) pixel sensors, 
called ``XRPIX'', for the next generation of X-ray astronomy satellites.
XRPIX is a monolithic active pixel sensor, fabricated using the SOI CMOS technology, and is 
equipped with the so-called ``Event-Driven readout'', 
which allows reading out only hit pixels by using the trigger circuit implemented in each pixel. 
The current version of XRPIX has lower spectral performance in the Event-Driven readout mode than in the Frame readout mode, 
which is due to the interference between the sensor layer and the circuit layer. 
The interference also lowers the gain. 
In order to suppress the interference, 
we developed a new device, ``XRPIX6E'' equipped with the Pinned Depleted Diode structure. 
A sufficiently highly-doped buried p-well is formed at the interface between the buried oxide layer and the sensor layer, and acts as a shield layer. 
XRPIX6E exhibits improved spectral performances both in the Event-Driven readout mode 
and in the Frame readout mode in comparison to previous devices. 
The energy resolutions in full width at half maximum at 6.4~keV are 236 $\pm$ 1 eV and 335 $\pm$ 4 eV in the Frame and Event-Driven readout modes, respectively. 
There are differences between the readout noise and the spectral performance in the two modes, 
which suggests that some mechanism still degrades the performance in the Event-Driven readout mode. 
\end{abstract}

\end{frontmatter}


\section{Introduction}
Charge-coupled devices (CCDs) are used as the standard imaging spectrometer 
in modern X-ray astronomy satellites. 
They have Fano limited spectroscopic performance with a readout noise level better than several electrons (rms) and 
fine imaging capability with a sensor size of $\sim25~\rm{mm} \times 25~\rm{mm}$ with a small pixel size ($\sim20~\rm{\mu m} \times 20~\rm{\mu m}$) (e.g.,~\cite{Koyama+07}). 
On the other hand, the time resolution of CCDs is $\sim$~1--10$~{\rm s}$, which is too slow to observe fast phenomena of compact objects such as neutron stars and blackholes. 
The slow readout causes photon pile-up problems resulting in serious degradation of the energy resolution.
Therefore, we have been developing a new type of X-ray pixel sensors, ``XRPIX'', aiming at a largely improved time resolution better than $\sim 10~{\rm \mu s}$~\cite{T.G.Tsuru}.
XRPIXs are active pixel sensors (APSs), processed with the silicon-on-insulator (SOI) CMOS technology~\cite{Y.Arai+11}. 
The SOI pixel sensor is monolithic using a low-resistivity Si bonded wafer for high-speed CMOS circuits (circuit layer), 
a SiO$_{2}$ insulator (Buried Oxide layer, or BOX layer), 
and a high-resistivity depleted Si layer for X-ray detection (sensor layer).
Each pixel has a trigger function as well as a low noise analogue readout circuit, 
which is the most important feature of the device. 
Reading out only triggering pixels allows obtaining good time resolution 
with a high throughput of $\sim 1~{\rm kHz}$, largely mitigating the photon pile-up problems. 
We refer to the readout mode using the trigger function as the ``Event-Driven readout mode". 
We can also read out all the pixels of XRPIX serially like a CCD without using the trigger function, 
which is referred to as the ``Frame readout mode". 

One of the current important developments is improvement of the spectral performance. 
The best performance so far in the Frame readout mode is obtained with XRPIX3b, 
whose equivalent noise charge (ENC) and energy resolution are 
$35~{\rm e^{-}}$ (rms) and $320~{\rm eV}$ at 6~keV, respectively~\cite{A.Takeda+15}. 
Takeda~et~al. (2015) also reported that the ENC can be reduced by increasing the gain~\cite{A.Takeda+15}. 
According to Miyoshi~et~al. (2017)~\cite{Miyoshi+17}, 
the gain is degraded by the coupling capacitance between the sensing area in the sensor layer  
and the pixel charge-sensitive amplifier (CSA) circuitry in the circuit layer. 
They found that adding an additional shield layer between the sensor layer and the circuit layer is effective 
in reducing the coupling and increasing the gain. 
Another problem is the spectral performance in the Event-Driven readout mode. 
Takeda~et~al. (2015)~\cite{A.Takeda+15} reported that spectral performance significantly degrades 
in the Event-Driven readout mode compared to the Frame readout mode. 
The degradation is caused by the cross-talk 
between the sensing area in the sensor layer and the digital trigger circuitry in the circuit layer. 
Ohmura~et~al. (2016)~\cite{S.Ohmura+16} demonstrated 
that the cross-talk can be reduced by introducing an additional shield layer between the two layers. 
Thus, both problems of the spectral performance would be solved by introducing an additional shield layer. 

Ohmura~et~al. (2016)~\cite{S.Ohmura+16} and Miyoshi~et~al. (2017)~\cite{Miyoshi+13} applied the Double SOI (DSOI) structure to realize the shield layer. 
The DSOI has the shield layer inside the BOX layer. 
We report the results from the XRPIXs with DSOI, XRPIX6D, in a separate paper [Takeda et al., 2018, in prep]. 
Kamehama~et~al. (2018)~\cite{Kamehama+18} developed a pinned depleted diode (PDD) structure 
and demonstrated that the spectral performance is significantly improved in the Frame readout mode. 
A sufficiently highly-doped buried p-well (BPW) is formed at the interface between the BOX and sensor layers, and acts as the shield layer. 
In this paper, we introduce newly developed XRPIX, XRPIX6E, with the PDD structure, 
and report the first spectral performance results, 
especially from the Event-Driven readout mode. 

\section{Device Description}
\begin{figure}[htbp]
\centering
\includegraphics[width=.95\columnwidth]{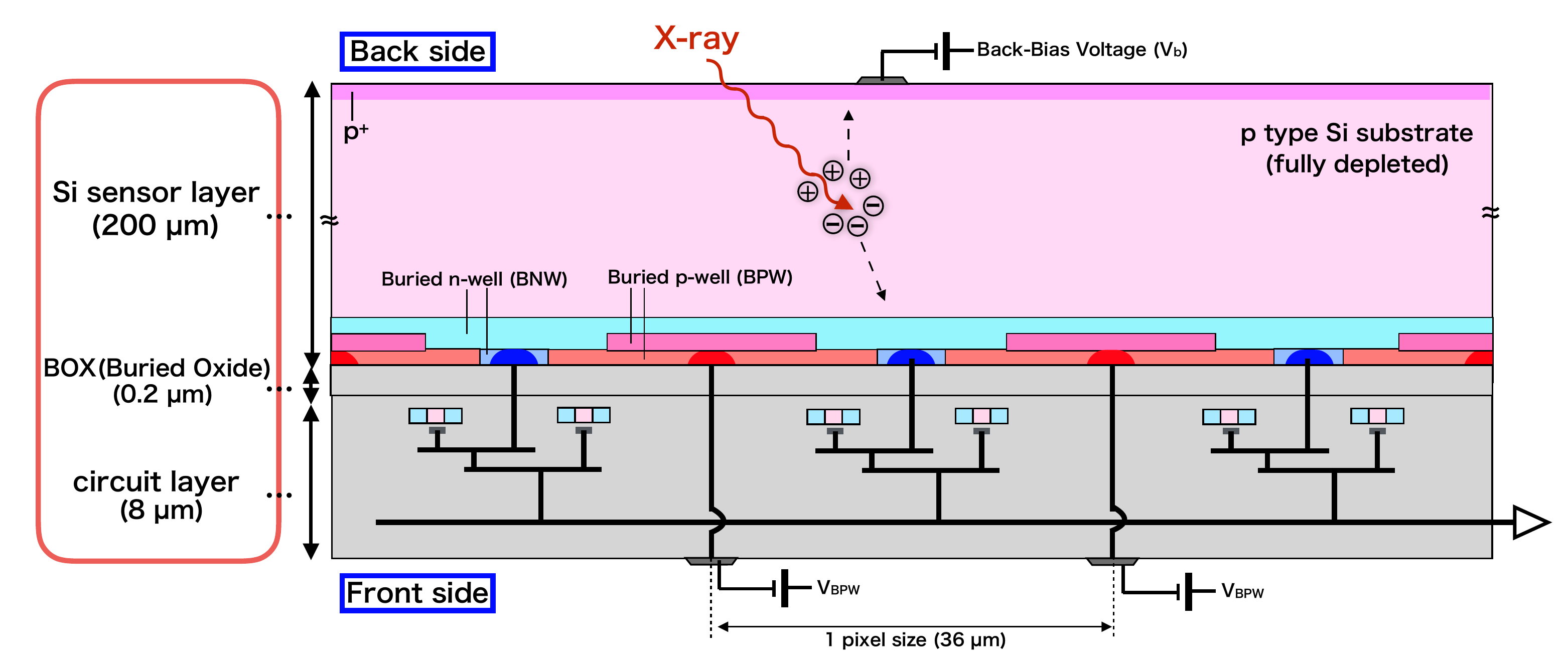}
\caption{Schematic cross-sectional view of XRPIX6E.}
\label{fig:XRPIX6E}
\end{figure}

The new device with the PDD structure, XRPIX6E, 
was fabricated by using a $0.2~\rm{\mu m}$ fully-depleted SOI CMOS pixel process supplied by Lapis Semiconductor Co. Ltd..
Figure~\ref{fig:XRPIX6E} shows a cross-sectional view of XRPIX6E. 
XRPIX6E has an identical PDD structure compared to SOIPIX-PDD of Kamehama et al. (2018)~\cite{Kamehama+18}.  
The sensor layer, having a thickness of $200~\rm{\mu m}$, is fabricated using a p-type floating zone (FZ) wafer with the nominal resistivity of $>25~{\rm k\Omega~cm}$. 
The charges generated by an X-ray are collected through the stepped buried n-well (BNW) by the readout 
and subsequently read by the CSA in the pixel circuit (Figure~\ref{fig:XRPIX6E}). 
The CSA is followed by a correlated double sampling circuit, which is also a part of the pixel circuit. The signal is then processed by the peripheral readout circuit consisting of a column Programmable Gain Amplifier (PGA) and an output buffer.

XRPIX6E and XRPIX6D have the same pixel and peripheral readout circuits including the layout. 
The details of the pixel and peripheral circuits will be published in Takeda et al. (2018, in prep). 
The pixel circuit of SOIPIX-PDD is different from that of XRPIX6E. 
In the case of SOIPIX-PDD, CDS circuit is implemented in the peripheral readout circuit. 
XRPIX6E has 2304 pixels arranged in $48 \times 48$. 
Each pixel has a size of $36~\rm{\mu m} \times 36~\rm{\mu m}$. 
On the front side of XRPIX is a circuit layer with a thickness of $8~\rm{\mu m}$, which absorbs soft X-rays 
and thus significantly reduces the quantum efficiency of the sensor. 
A hole ($3.5~\rm{mm} \times 3.5~\rm{mm}$) is drilled in the back side of the ceramic package so that the device is sensitive to soft X-rays with back-side illumination.

\section{Experiments and Results}
We used a readout system consisting of a printed circuit board (sub-Board) on which an XRPIX6E and a second stage amplifier with a gain of 1.8 are mounted,  
and a SEABAS (Soi EvAluation BoArd with Sitcp) board~\cite{T.Uchida}. 
Throughout the experiments in this paper, 
we applied a back-bias voltage of $V_{b}=-200~\rm{V}$ to the device so that the full-depletion of the sensor layer is realized. 
The voltage applied to the BPW ($V_{\rm BPW}$) was fixed at $-2~\rm{V}$. 
We set gain to unity on the PGA in the peripheral readout circuit of the device and 
the amplifier gain to 1.8 on the instrumentation amplifier assembled on the sub-Board [Takeda et al., 2018, in prep]. 
XRPIX6E was cooled to $-60^{\circ}{\rm C}$ with a thermostatic chamber in order to reduce the pixel dark current. 
The X-ray performance was evaluated using the radioisotopes $^{57}$Co and $^{241}$Am. 

We read out $8 \times 8$ pixels located at the center of the device 
following the sequences in the Frame and Event-Driven readout modes described 
in the papers by Ryu~et~al. (2011)~\cite{Ryu+11} and Takeda et~al. (2013)~\cite{A.Takeda+13}, respectively. 
We analysed the data by using the method described 
by Ryu et~al. (2011)~\cite{Ryu+11} and Nakashima et~al. (2012)~\cite{Nakashima+12}. 
We identified a pixel whose pulse height exceeds a predefined threshold called the event threshold in the Frame readout mode, 
whereas we identified an X-ray hit when voltage of the in-pixel comparator circuits exceeds 
a predefined threshold called the trigger threshold voltage in the Event-Driven readout mode. 
In both modes, we examined the pulse heights of 8 pixels adjacent to the identified or triggering pixel 
to see whether or not any of them exceeds the threshold for split signals (split threshold). 
Each X-ray event is classified into one of the following types: 
``single pixel'', ``double pixel'', ``triple pixel'' and ``other'', according to the pattern of the 8 pixels. 
In the following, we only used single pixel events for simplicity.

\subsection{Gain and its Uniformity of $8\times8$ pixels}\label{Gain and Uniformity}
\begin{figure}[htbp]
\centering
\includegraphics[width=.75\columnwidth]{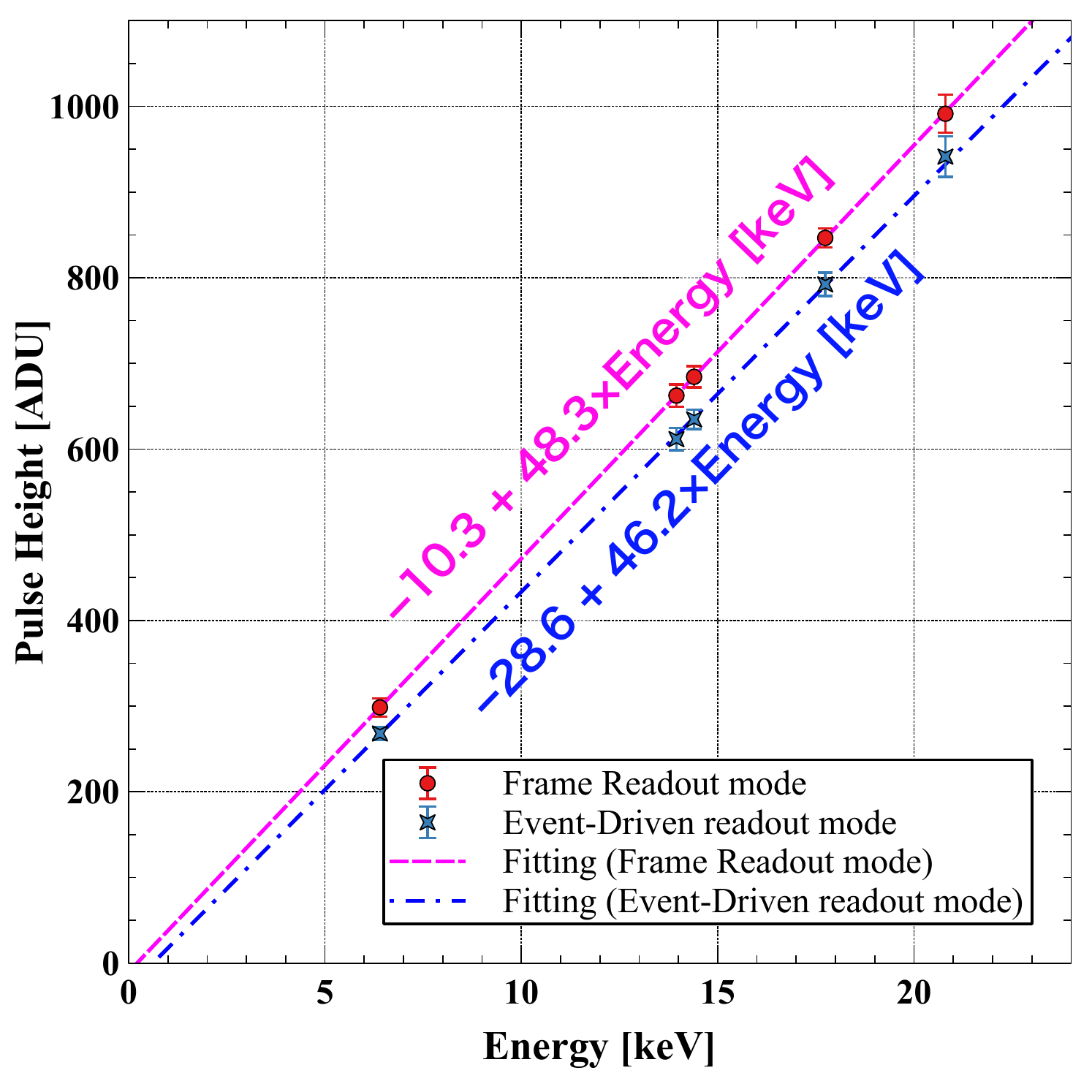}
\caption{Peak amplitude as a function of X-ray energy for the Frame readout mode (magenta; dash line), 
and in the Event-Driven readout mode (blue; dash-dot line), 
where 1 ADU corresponds to $488~\rm{\mu V}$. 
The error bars indicate $1\sigma$ statistical uncertainties. 
The best-fit linear functions are also shown. 
Note that the pulse heights in this figure are amplified by the second stage amplifier with a gain of 1.8.}
\label{Fig: ADU_Energy_plot_20180321}
\end{figure}

Figure \ref{Fig: ADU_Energy_plot_20180321} shows the pulse height as a function of X-ray energy. 
The data points at $6.4~\rm{keV}$ and $14.4~\rm{keV}$ were obtained with $^{57}$Co, and 
those at $13.95~\rm{keV}$, $17.75~\rm{keV}$, and $20.8~\rm{keV}$ were obtained with $^{241}$Am.
A good linearity was obtained both in the Frame and in the Event-Driven readout modes 
with the gains of $(47.8 \pm 1.2)~{\rm \mu V/e^{-}}$ and $(45.7 \pm 0.5)~{\rm \mu V/e^{-}}$ for the Frame and Event-Driven readout modes, respectively. 
These values were calculated using $1~{\rm ADU} = 488~{\rm \mu V}$, and a mean ionization energy in silicon, 3.65~{\rm eV}.

As seen in Figure \ref{Fig: ADU_Energy_plot_20180321}, the relations between X-ray energy and output pulse 
height obtained in the two readout modes are almost consistent with each other. 
Our previous devices, XRPIX2b and XRPIX3b, had large offsets in output when operated in 
the Event-Driven mode~\cite{S.Ohmura+16, A.Takeda+14}, which can be attributed to the interference between the sensor and circuit layers. 
The significantly reduced offset, therefore, indicates that the interference is largely suppressed in XRPIX6E, 
and we came to a conclusion that the highly doped BPW of the PDD structure acts effectively as a shield.

\begin{figure}[htbp]
\centering
\includegraphics[width=.95\columnwidth]{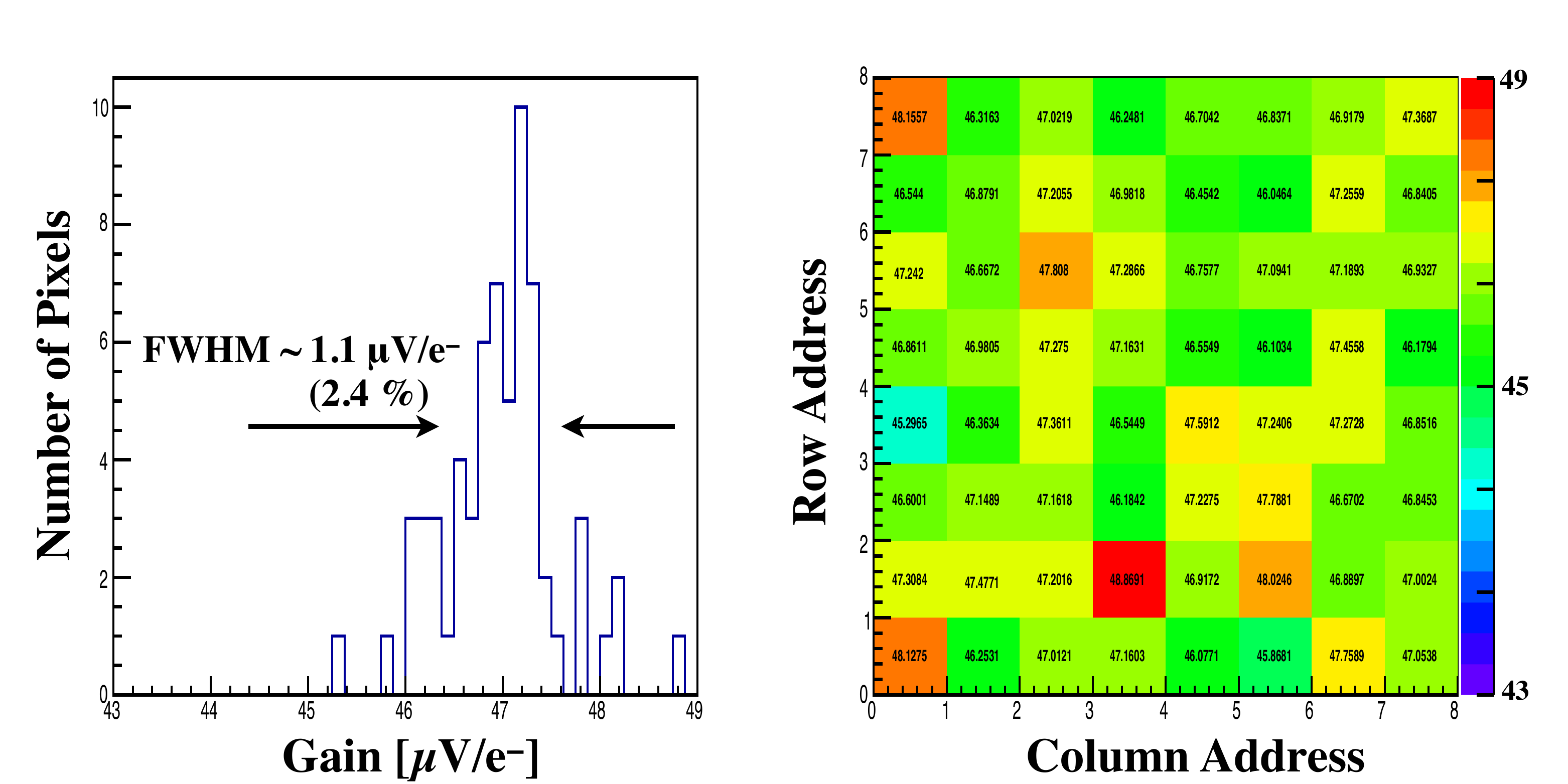}
\caption{Gain variation of the $8 \times 8$ pixels plotted as a histogram (left) and visualized as a map (right).}
\label{Fig: gain distribution}
\end{figure}

Unlike X-ray CCDs, APSs are usually subject to gain non-uniformity. 
Figure~\ref{Fig: gain distribution} shows the gain distribution of the 8$\times$8 pixels obtained in the Frame readout mode. 
The pixel-to-pixel gain variation is 2.4\% in full width at half maximum (FWHM). 

\subsection{Spectral Performances}
\begin{figure}[htbp]
	\begin{tabular}{c}
		\begin{minipage}[b]{0.5\hsize}
			\centering
			\includegraphics[width=.95\columnwidth]{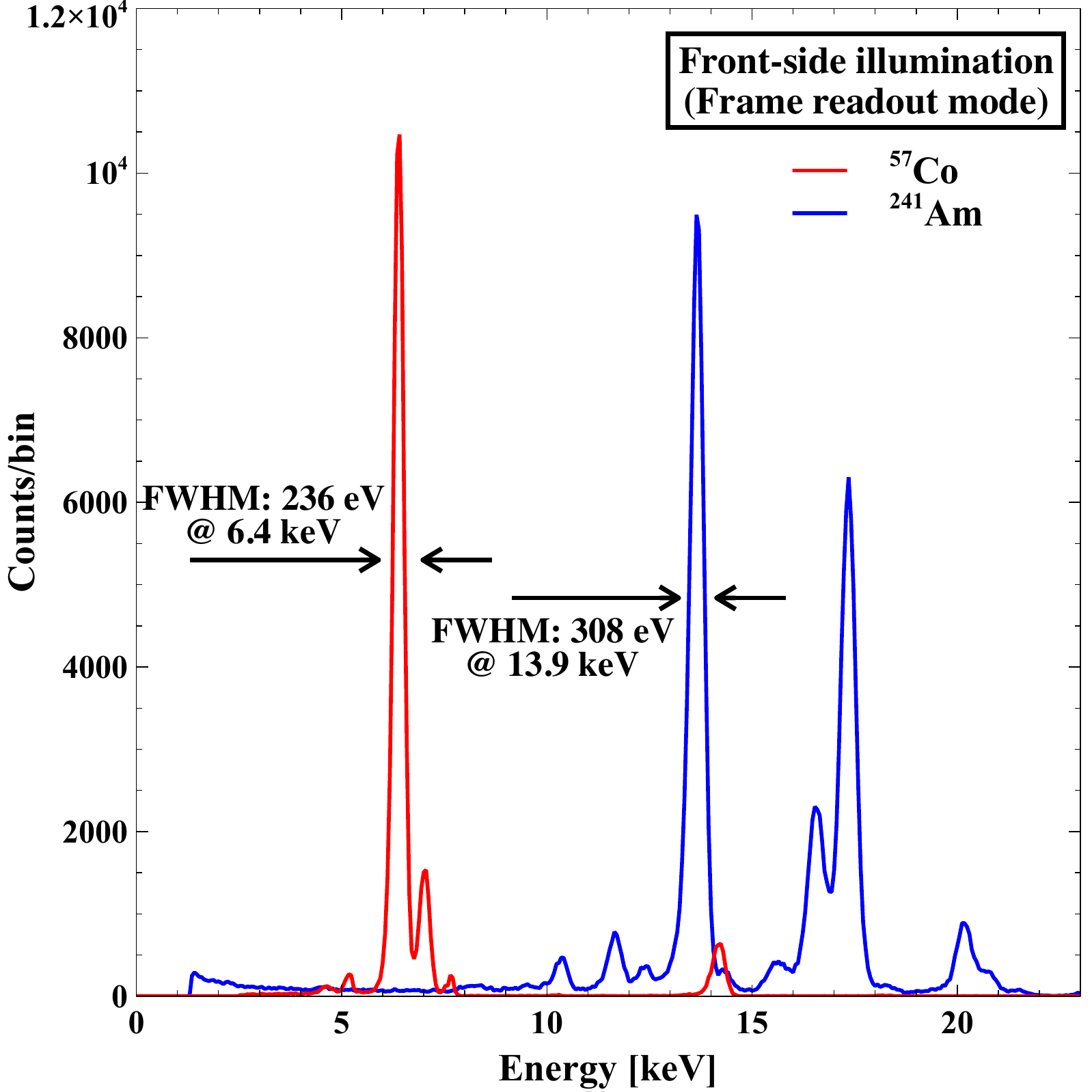}
		\end{minipage}
		\begin{minipage}[b]{0.5\hsize}
			\centering
			\includegraphics[width=.95\columnwidth]{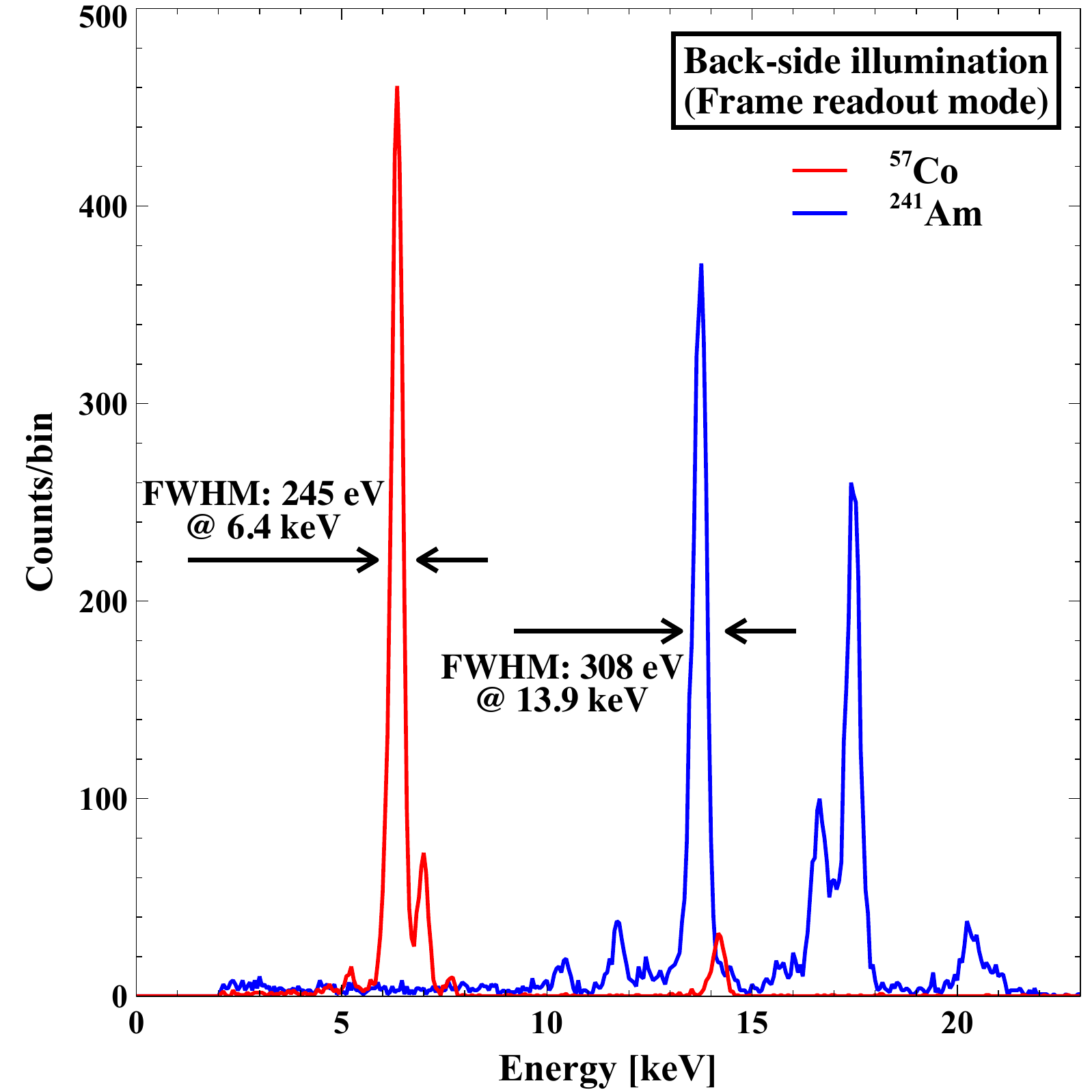}
		\end{minipage}
	\end{tabular}

	\begin{tabular}{c}
		\begin{minipage}[b]{0.5\hsize}
			\centering
			\includegraphics[width=.95\columnwidth]{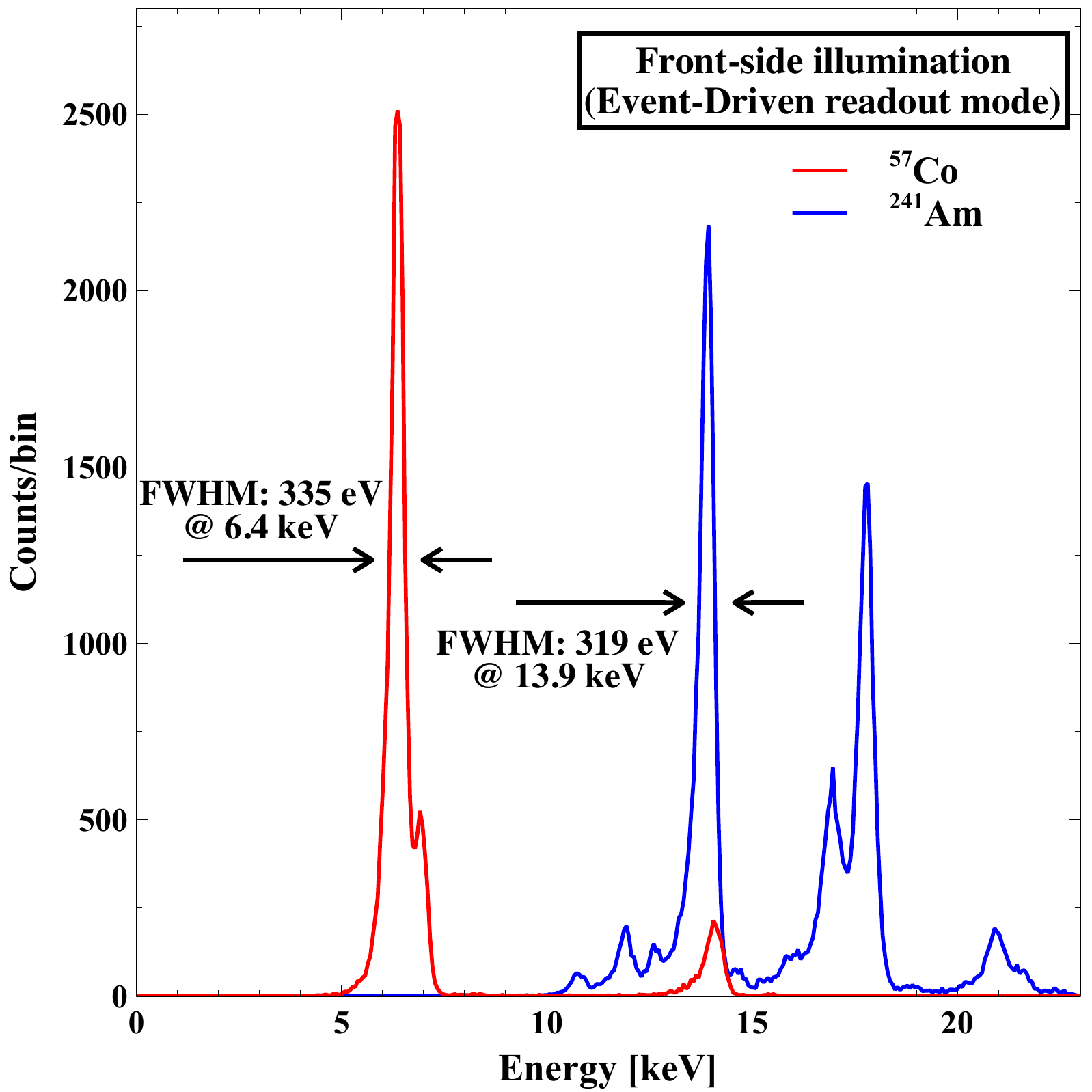}
		\end{minipage}
		\begin{minipage}[b]{0.5\hsize}
			\centering
			\includegraphics[width=.95\columnwidth]{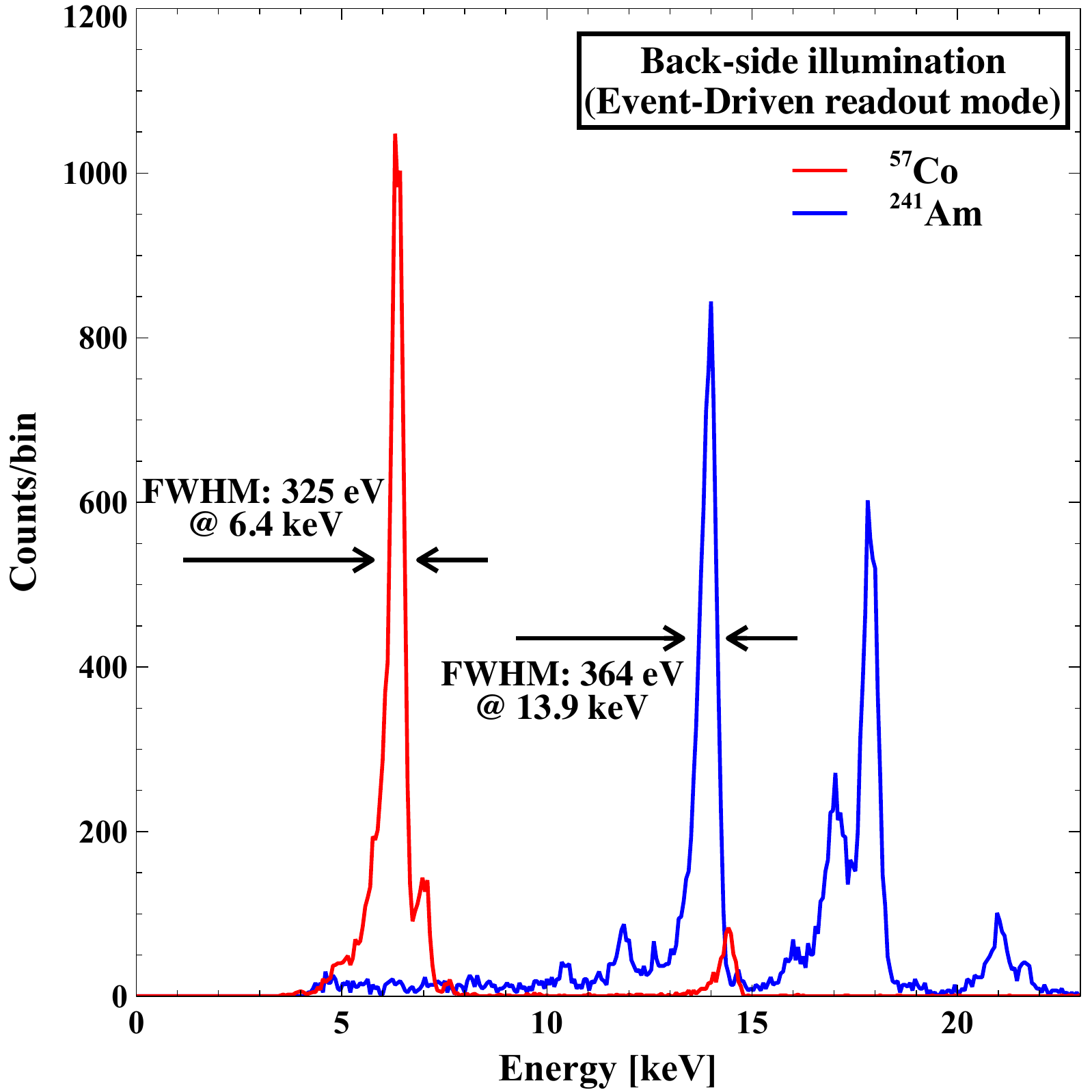}
		\end{minipage}
	\end{tabular}

\caption{X-ray spectra obtained with XRPIX6E. 
(Top left) Spectra obtained Frame readout mode with the front-side illumination, 
(top right) Frame readout mode with the back-side illumination, (bottom left) Event-driven readout mode with the front-side illumination, and (bottom right) Event-driven readout mode with the back-side illumination.}
\label{Fig: XR6E_gain_revise}
\end{figure}

Figures~\ref{Fig: XR6E_gain_revise} shows spectra of single-pixel events in the Frame and Event-Driven readout modes. 
The correction of the pixel-to-pixel gain variation is applied to the spectra, 
based on the gain map obtained in Section~\ref{Gain and Uniformity}. 
The energy resolutions at 6.4~keV and 13.9~keV are summarized in Table~\ref{Tab: Energy Resolution}. 
The ENCs are $22.0 \pm 0.1~{\rm e}^-$ (rms) in the Frame readout mode with the front-side illumination
and $35.6 \pm 0.1~{\rm e}^-$ (rms) in the Event-Driven readout modes with the front-side illumination. 
We estimated the ENCs by evaluating one sigma widths of the pedestal peak and converting them using the gains as
\begin{eqnarray}
{\rm ENC} = \sigma_{\rm ped}/G/\varepsilon_{\rm  Si}, 
\end{eqnarray}
where $\sigma_{\rm ped}$ is the pedestal peak width, $G$ is the gain, and $\varepsilon_{\rm  Si} = 3.6~{\rm eV}$ is the mean energy to produce an electron-hole pair in Si.

\begin{table}
 \begin{center}
  \caption{Energy Resolution}
   \begin{tabular}{lccc}
   \hline
   Mode & Illumination & 6.4~keV & 13.9~keV \\ 
        &              & (FWHM)  & (FWHM) \\ 
   \hline
   Frame & Front side  & 236~$\pm$~1 eV  & 308~$\pm$~3 eV\\
         & Back side   & 245~$\pm$~8 eV  & 308~$\pm$~20 eV \\
   Event & Front side  & 335~$\pm$~4 eV  & 319~$\pm$~4 eV \\
         & Back side   & 325~$\pm$~7 eV  & 364~$\pm$~40 eV \\
   \hline
   \end{tabular}
 \label{Tab: Energy Resolution}
 \end{center}
\end{table}

\section{Discussion}
\begin{figure}[htbp]
\centering
\includegraphics[width=.75\columnwidth]{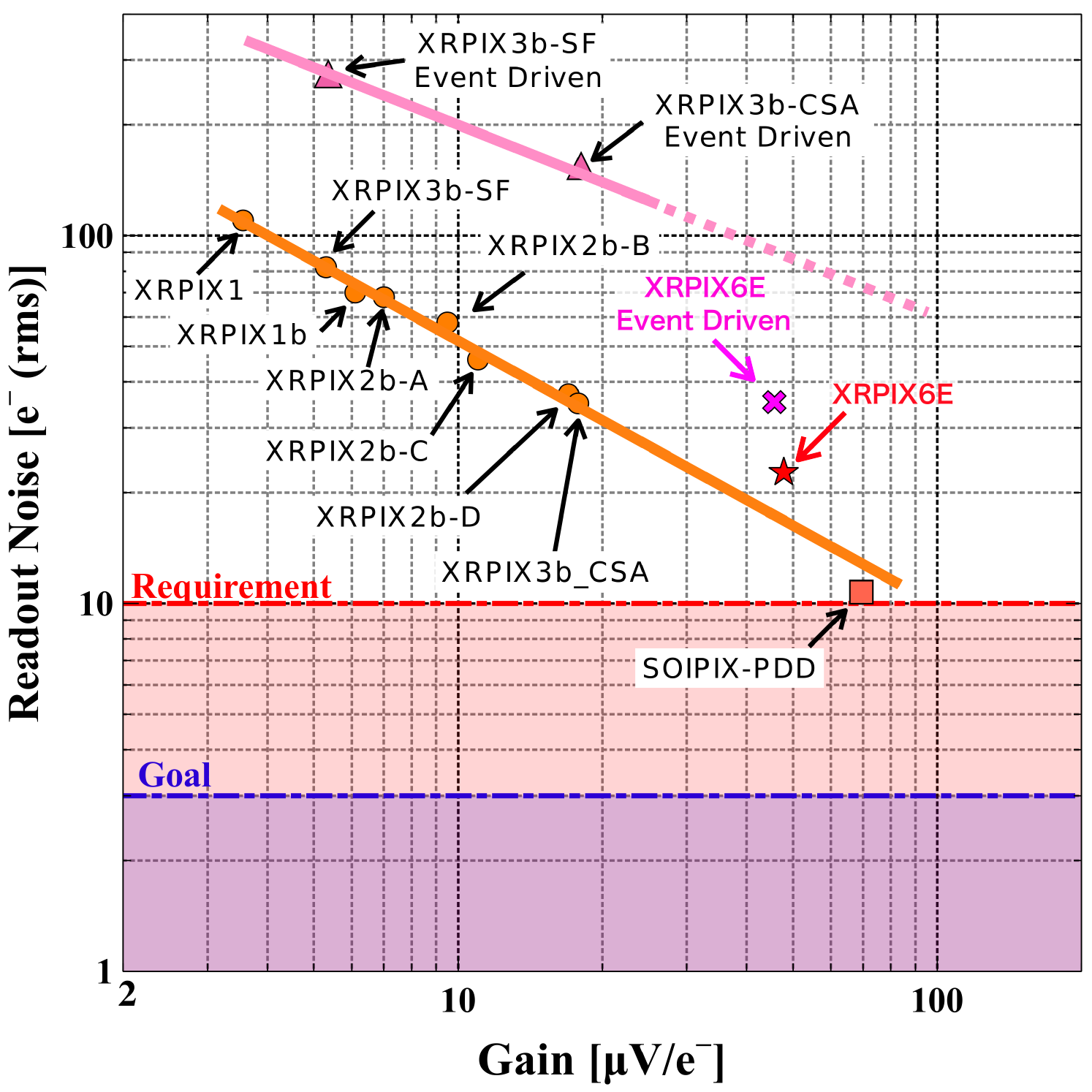}
\caption{Relation between the gain and the readout noise. 
The data points without the label of ``Event Driven'' are obtained 
in the Frame readout mode~\cite{A.Takeda+15, S.Ohmura+16, Kamehama+18}. 
The two solid lines are eye guides for the correlation.}
\label{Fig: takeda_plot}
\end{figure}

In this section, we discuss the spectral performance of XRPIX6E in comparison with other devices of XRPIX series. 
As shown in Section~\ref{Gain and Uniformity}, the gain of XRPIX6E, $46$--$48~{\rm \mu V/e^{-}}$, is significantly higher than 
$17.8~{\rm \mu V/e^{-}}$ of XRPIX3b~\cite{A.Takeda+15}. 
The two devices are equipped with the CSAs with the same design. 
Takeda~et~al. (2018, in prep) will report that the gain is degraded by the coupling capacitance between the sensing area in the sensor layer 
and the pixel CSA circuitry in the circuit, which makes the feedback capacitance of the CSA effectively smaller.  
The significantly increased gain, therefore, indicates that the parasitic capacitance between the sensing area and the CSA decreased due to the highly doped BPW of the 
PDD structure acting as a shield.
 On the other hand, the gain of XRPIX6E is $\sim 2/3$ of that for the SOIPIX-PDD~\cite{Kamehama+18}, 
which is due to the difference in the design of the CSAs between the two devices. 
The measured pixel-to-pixel gain variations of 2.4\% is comparable with those of the previous XRPIX devices: 3.3\% for XRPIX2 and 1.8\% for XRPIX5b~\cite{Nakashima+13, Hayashi+18}. 
Figure~\ref{Fig: takeda_plot} shows the readout noise as a function of gain of XRPIX series, 
including XRPIX6E and SOIPIX-PDD. 
The readout noise of XRPIX6E operated in the Frame mode is consistent with the expectation from the other devices of the XRPIX series, 
which indicates that the reduction in the readout noise is primarily due to the increase in the gain.

In the Event-Driven readout mode, the readout noise of XRPIX6E is significantly lower than the extrapolation of the data points from XRPIX3b 
(Figure~\ref{Fig: takeda_plot}), indicating that the increase of the gain alone cannot account for the improvement of the noise performance. 
The most plausible explanation is that the better performance is achieved thanks to the significantly reduced interference between the sensor and circuit layers by the PDD structure. 
XRPIX6E has the energy resolution of $\sim330~{\rm eV}$ in FWHM at $6.4~{\rm keV}$ in the Event-Driven readout mode, 
which is the best in the XRPIX series.

On the other hand, the readout noise in the Event-Driven readout mode is still higher than that in the Frame readout mode. 
The two modes have slightly different gains and offsets in the output voltage as a function of X-ray energy. 
This result shows that some mechanism still degrades the performance in the Event-Driven readout mode. 
Identifying the cause and improving the performance are major tasks for the future.

Finally, we discuss the charge collection efficiency. 
A large low energy tail structure is observed in XRPIX5b~\cite{Hayashi+18}.
XRPIX6E has a smaller low energy tail structure in the spectrum than shown by XRPIX5b.
XRPIX6E has similar spectral performances with the back-side illumination to those with the front-side illumination, 
at least in the energy band of $\sim$~6--20~keV (see Figures~\ref{Fig: XR6E_gain_revise}). 
These results indicate that the PDD structure improves the charge collection efficiency 
as discussed by Kamehama~et~al. (2018)~\cite{Kamehama+18}.

\section{Conclusions}
XRPIX6E with the Pinned Depleted Diode (PDD) structure is presented in this paper. 
The XRPIX6E design has successfully suppressed the interference between the sense-node and the circuit by introducing the PDD structure.
 The PDD structure improves the spectral performance in the Event-Driven readout mode in addition to the Frame readout mode, 
and the charge collection efficiency. 
The energy resolution of 335~$\pm$~4 eV in FWHM at 6.4~keV in the Event-Driven readout mode is the best in XRPIX series. 
On the other hand, the spectral performance in the Event-Driven readout mode is still degraded 
in comparison with the Frame readout mode. 


\section*{Acknowledgements}
We acknowledge the valuable advice and great work by the personnel of LAPIS Semiconductor Co., Ltd. 
This study was supported by the Japan Society for the Promotion of Science (JSPS) KAKENHI Grant-in-Aid for Scientific Research on Innovative Areas 25109002 (Y.A.), 25109003 (S.K.), 25109004 (T.G.T. and T.T.), 20365505 (T.K.), 23740199 (T.K.), 18740110 (T.K.), Grant-in-Aid for Young Scientists (B) 15K17648 (A.T.), Grant-in-Aid for Challenging Exploratory Research 26610047 (T.G.T.) and Grant-in-Aid for JSPS Fellows 15J01842 (H.M.). 
This study was also supported by the VLSI Design and Education Center (VDEC), the University of Tokyo in collaboration with Cadence Design Systems, Inc., and Mentor Graphics, Inc.


\begin{thebibliography}{99}

\bibitem{Koyama+07}
K. Koyama, H. Tsunemi, T. Dotani et al., 
X-Ray Imaging Spectrometer (XIS) on Board Suzaku, 
Publications of the Astronomical Society of Japan, 59 (2007) S23. 

\bibitem{T.G.Tsuru}
T. G. Tsuru, H. Matsumura, A. Takeda et al.,  
Development and performance of Kyoto's x-ray astronomical SOI pixel (SOIPIX) sensor, 
in Proc. SPIE, 9144, (2014) 914412. 

\bibitem{Y.Arai+11}
Y. Arai, T. Miyoshi, Y. Unno et al., 
Development of SOI Pixel Process Technology, 
Nuclear Instruments and Methods in Physics Research Section A, 636 (2011) S31. 

\bibitem{A.Takeda+15}
A. Takeda, T. G. Tsuru, T. Tanaka et al., 
Improvement of spectroscopic performance using a charge-sensitive amplifier circuit for an X-ray astronomical SOI pixel detector, 
Journal of Instrumentation, 10 (2015) C06005.

\bibitem{Miyoshi+17}
T. Miyoshi, Y. Arai, Y. Fujita et al., 
Front-end electronics of double SOI X-ray imaging sensors, 
Journal of Instrumentation, 12 (2017) C02004

\bibitem{S.Ohmura+16}
S. Ohmura, T. G. Tsuru, T. Tanaka et al., 
Reduction of cross-talks between circuit and sensor layer in the Kyoto's X-ray astronomy SOI pixel sensors with Double-SOI wafer, 
Nuclear Instruments and Methods in Physics Research Section A, 831 (2016) 61

\bibitem{Miyoshi+13}
T. Miyoshi, Y. Arai, T. Chiba et al., 
Monolithic pixel detectors with ${\rm 0.2\mu m}$ FD-SOI pixel process technology
Nuclear Instruments and Methods in Physics Research Section A, 732 (2013) 530


\bibitem{Kamehama+18}
H. Kamehama, S. Kawahito, S. Shrestha et al., 
A Low-Noise X-ray Astronomical Silicon-On-Insulator Pixel Detector Using a Pinned Depleted Diode Structure, 
Sensors, 18 (2018) 27

\bibitem{T.Uchida}
T. Uchida, 
Hardware-Based TCP Processor for Gigabit Ethernet, 
IEEE Transactions on Nuclear Science, 55 (2008) 1631.

\bibitem{Ryu+11}
S. G. Ryu, T. G. Tsuru, S. Nakashima et al., 
First Performance Evaluation of an X-Ray SOI Pixel Sensor for Imaging Spectroscopy and Intra-Pixel Trigger, 
IEEE Transactions on Nuclear Science, 58 (2011) 2528.
\
\bibitem{A.Takeda+13}
A. Takeda, Y. Arai, S. G. Ryu et al., 
Design and Evaluation of an SOI Pixel Sensor for Trigger-Driven X-Ray Readout, 
IEEE Transactions on Nuclear Science, 60 (2013) 586

\bibitem{Nakashima+12}
S. Nakashima, S. G. Ryu, T. G. Tsuru et al., 
Progress in Development of Monolithic Active Pixel Detector for X-ray Astronomy with SOI CMOS Technology, 
Physics Procedia, 37 (2012) 1373.

\bibitem{A.Takeda+14}
A. Takeda, T. G. Tsuru, T. Tanaka et al., 
Development and Evaluation of Event-Driven SOI Pixel Detector for X-ray Astronomy, 
Proceedings of Science (TIPP2014), 213 (2014) id138

\bibitem{Nakashima+13}
S. Nakashima, S. G. Ryu, T. Tanaka et al., 
Development and characterization of the latest X-ray SOI pixel sensor for a future astronomical mission, 
Nuclear Instruments and Methods in Physics Research Section A, 731 (2013) 74

\bibitem{Hayashi+18}
H. Hayashi, T. G. Tsuru, T. Tanaka et al., 
Evaluation of Kyoto's Event-Driven X-ray Astronomical SOI Pixel Sensor with a Large Imaging Area, 
Nuclear Instruments and Methods in Physics Research Section A, submitted to this issue. 


\end{thebibliography}
\end{document}